\documentclass[twocolumn]{revtex4}
\usepackage{graphicx}
\usepackage{hyperref}

\def\mathbi#1{\textbf{\em #1}}
\newcommand{\Rm}{R_0}
\newcommand{\pv}{p_{\rm v}}
\newcommand{\del}{\mbox{\boldmath$\nabla$}}
\newcommand{\g}{{\mathbi g}}
\renewcommand{\a}{{\mathbi a}}
\newcommand{\Dp}{\Delta p}

\begin{document}

\title{Bubble-Jets in Gravitational Fields (APS DFD Video V060)}
\author{D. Obreschkow}
\author{M. Tinguely}
\author{N. Dorsaz}
\author{P. Kobel}
\author{A. de Bosset}
\author{M. Farhat}
\affiliation{Ecole Polytechnique F\'{e}d\'{e}rale de Lausanne, LMH, 1007 Lausanne, Switzerland}
\date{\today}
\begin{abstract}
We show visualizations of the gravity-induced jets formed by spherical bubbles collapsing in liquids subjected to normal gravity, micro-gravity, and hyper-gravity. These observations demonstrate that gravity can have a significant effect on cavitation bubbles. An analysis of the gravity-induced jets, detailed in Ref.~\cite{Obreschkow2011b}, uncovers a scaling law between the size of bubble-induced jets and the non-dimensional parameter $\zeta\equiv|\del p|\Rm/\Dp$, where $\Rm$ is the maximal bubble radius and $\Delta p$ is the driving pressure. This scaling law applies to any jet formed in any uniform pressure gradient.
\end{abstract}
\maketitle


Jets produced by collapsing cavitation bubbles play a key role in modern technologies \citep{Leighton2010} and erosion \citep{Ohl2006}. This paper contributes to the study of such jets through precise visualizations and universal results of jets formed by bubbles collapsing in different gravitation fields.

Our study relies on an experiment producing the most spherical cavitation bubbles today. The bubbles grow inside a liquid from a point-plasma generated by a nanosecond laser pulse. Unlike in previous studies, the laser is focussed by a parabolic mirror, resulting in a plasma of unprecedented symmetry. The arising bubbles are sufficiently spherical that the hydrostatic pressure gradient caused by gravity becomes the dominant source of asymmetry in the collapse and rebound of the cavitation bubbles. To avoid this natural source of asymmetry, the whole experiment is therefore performed in micro-gravity conditions (ESA, 52nd parabolic flight campaign).

\begin{figure}[h]
	\includegraphics[width=\columnwidth]{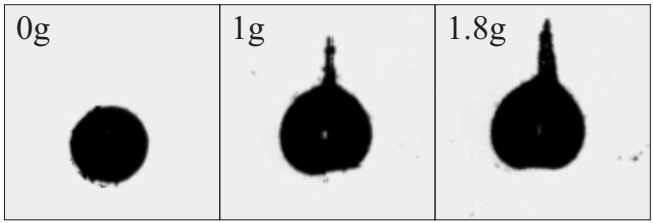}
	\caption{Comparison of three initially spherical cavitation bubbles during their rebound phase after collapse. The bubbles evolved in different gravitational accelerations $\a\in\{0\g,1\g,1.8\g\}$ implying different pressure gradients $\del p=\rho\a$. The bubble radii are $\Rm\in\{4.5,4.3,4.5\}\rm\,mm$ and the driving pressures are $\Delta p\in\{26.0,7.2,9.0\}\rm\,kPa$.}
	\label{fig}
\end{figure}

Cavitation bubbles were observed in micro-gravity ($0g$) and in normal gravity ($1g$) to hyper-gravity ($1.8g$). The \href{http://bubbles.epfl.ch/files/content/sites/bubbles/files/APS_DFD_V060_2011.mp4}{Video} featured with this paper shows the full evolution of three initially spherical bubbles subjected these three gravitational fields. A single frame from this video is represented in Fig.~\ref{fig} and shows the three bubbles during their rebound phase. This figure demonstrates jet formation in the weak uniform pressure gradient of gravity.

We systematically analyzed the jets of collapsing spherical bubbles, while varying four experimental parameters: the gravitational acceleration $\a$ ($0\g-1.8\g$), the maximal bubble radius $\Rm$ ($1-7~\rm mm$), the mean driving pressure $\Dp$ ($8-80\rm~kPa$) defined as the difference between the mean static pressure $p$ and the vapor pressure $\pv$, and the viscosity $\eta$ ($1-30\rm~mPa~s$). The latter was varied by choosing specific water-glycerine mixtures for the liquid. A statistical analysis of the data unveiled a proportionality relation between the normalized jet volume of the cavitation bubbles in a uniform pressure gradient and the non-dimensional parameter (plots in Ref.~\cite{Obreschkow2011b})
\begin{equation}
	\zeta\equiv|\del p|\Rm/\Dp.
\end{equation}

We derived a physical model \cite{Obreschkow2011b} based on energy and momentum condervation \cite{Blake1988}, which explains the proportionality relation between the normalized jet volume and $\zeta$. This model also explains the negligible dependence of the jet volume on $\eta$, as well as on surface tension $\sigma$.

In summary, this research shows unprecedented visualizations of gravity-driven jets of cavitation bubbles. By systematically analyzing and modelling these jets, we uncovered a universal scaling law, which establishes a general link between a non-dimensional parameter $\zeta$ and the jet volume produced by spherical bubbles collapsing in a uniform pressure gradient $\del p$. Since such a uniform gradient is the linear term in the Taylor expansion of any smooth pressure field, our result extends, to first order, to bubbles in any other pressure field.

Supported by the European Space Agency ESA and the Swiss NSF (200020-116641, PBELP2-130895).

\end{document}